\documentclass[preprint,proceedings]{rmaa}

\usepackage{paralist}

\usepackage{color}

\def\ts{\thinspace}
 
\def\gapprox{$_>\atop{^\sim}$} \def\lapprox{$_<\atop{^\sim}$}

\newdimen\sa  \def\sd{\sa=.1em \ifmmode $\rlap{.}$''$\kern -\sa$
                               \else \rlap{.}$''$\kern -\sa\fi}

\def\m31{M{\ts}31} \def\mm32{M{\ts}32} 

\def\0{\phantom{0}}
\def\ba{\kern -1pt}

\def\dot{\0\ts\raise 0.2em\hbox{{\dots}}}

\SetYear{2005}
\SetConfTitle{The Ninth Texas-Mexico Conference on Astrophysics}

\title{Secular Evolution in Disk Galaxies:\\ 
       The Growth of Pseudobulges and\\
       Problems for Cold Dark Matter Galaxy Formation} 

\author{
        John Kormendy\altaffilmark{1}
    and David B.~Fisher\altaffilmark{1}}

\fulladdresses{
\item Department of Astronomy, RLM 15.308, University of Texas,
      Austin, TX 78712, USA.}

\altaffiltext{1}{Department of Astronomy, University of Texas, Austin.}

\shortauthor{Kormendy \& Fisher}
\shorttitle{Internal Secular Evolution in Galaxies}

\fulladdresses{
\item David B.~Fisher and John Kormendy,  Department of Astronomy, 
      University of Texas at Austin, 1 University Station,
      Austin, TX 78712, USA (twitch@astro.as.utexas.edu, 
      kormendy@stro.as.utexas.edu).

}

\listofauthors{John Kormendy \& David B. Fisher}
\indexauthor{Kormendy, J.}
\indexauthor{Fisher, D. B.}

\abstract{
      We review internal secular evolution in galaxy disks -- the
fundamental process by which isolated disks evolve.  We concentrate 
on the buildup of dense central features that look like classical, 
merger-built bulges but that were made slowly out of disk gas.  We 
call these pseudobulges.  As an existence proof, we review how bars 
rearrange disk gas into outer rings, inner rings, and gas dumped
into the center.  In simulations, this gas reaches high densities, and in
the observations, many SB and oval galaxies show central concentrations
of gas.  Associated star formation rates imply plausible pseudobulge
growth times of a few billion years.

      If secular processes built dense centers that
masquerade as bulges, can we distinguish them from merger-built bulges? 
Observations show that pseudobulges retain a memory of their disky
origin.  They have one or more characteristics of disks: (1) flatter 
shapes than those of classical bulges, (2) larger ratios of ordered to 
random velocities, (3) smaller velocity dispersions, (4) nuclear bars or
spiral structure, (5) boxy structure when seen edge-on, (6) nearly 
exponential brightness profiles, and (7) starbursts.  These features occur
preferentially in barred and oval galaxies in which secular
evolution should be rapid.  So the cleanest examples of pseudobulges are
recognizable.  Thus observations and theory contribute to a new picture
of galaxy evolution that complements hierarchical clustering and merging.

      However, an important problem with cold dark matter galaxy
formation gets more acute.  How can hierarchical clustering produce so many
pure disk galaxies with no evidence for merger-built bulges?
}

\resumen{\vskip 200pt}

\addkeyword{galaxies: evolution}
\addkeyword{galaxies: formation}
\addkeyword{galaxies: kinematics and dynamics}
\addkeyword{galaxies: nuclei}

\begin{document}

\maketitle

\pretolerance=10000   \tolerance=10000

\section{Introduction}

      Galactic evolution is in transition from the early Universe dominated by
hierarchical clustering to a future dominated by internal secular processes. 
These result from interactions involving collective phenomena such as bars, 
oval disks, spiral structure, and triaxial dark halos.  This paper summarizes
and updates reviews by Kormendy (1993) and especially by Kormendy \& Kennicutt
(2004, hereafter KK04).  

\section{The Fundamental Way that Disks Evolve is by Spreading}

      A general principle\footnote{Exceptions exist but are rare and somewhat
contrived.} of the evolution of
self-gravitating systems is that it is energetically favorable to spread -- 
to shrink the inner parts by expanding the outer parts.  The easiest way to 
see this depends on whether the system is dominated by rotation or by random
motions.

\subsection{If Dynamical Support Is By Random Motions}

\def\ts{\thinspace}

      Then the argument (Lynden-Bell \& Wood 1968;
Binney\ts\&{\ts}Tremaine\ts1987){\ts}is{\ts}based{\ts}on{\ts}the{\ts}fundamental
point that the specific heat of a self-gravitating
system is negative.  Consider an equilibrium system of $N$ particles of mass
$m$, radius~$r$, and three-dimensional velocity dispersion~$v$.~The virial
theorem says that 2{\ts}KE + PE = 0, where the kinetic energy KE = $Nmv^2/2$
and the potential energy PE = $-G(Nm)^2/r$ define $v$~and~$r$. The total energy
of a bound system, $E \equiv$ KE $+$ PE = $-$KE, is negative.  But temperature
$T$ corresponds to internal velocity as $mv^2/2 = 3kT/2$.  So the specific heat 
$C \equiv dE/dT \propto d(-Nmv^2/2)/d(v^2)$ is also negative.  In the above, $G$ is
the gravitational constant and $k$ is Boltzmann's~constant.

      The system is supported by heat, so evolution is by heat transport.
If the center of the system is hotter than the periphery, then heat tends to flow 
outward.  The inner parts shrink and get still hotter.  This promotes further
heat flow. The outer parts receive heat; they expand and cool.  Whether the
system evolves on an interesting timescale depends on whether there is an
effective heat-transport mechanism.  For example, many globular clusters evolve
quickly by two-body relaxation and undergo core collapse.  Giant 
elliptical galaxies -- which otherwise would evolve similarly -- cannot do so
because their relaxation times are much longer than the age of the Universe.

\subsection{If Dynamical Support Is By Rotation}

      Tremaine (1989) provides a transparent summary of an argument due to
Lynden-Bell \& Kalnajs (1972) and to Lynden-Bell \& Pringle (1974).  A disk is
supported by rotation, so evolution is by angular momentum transport.  The
``goal'' is to minimize the total energy at fixed total angular momentum.  
A rotationally supported ring at radius $r$ in a fixed potential $\Phi(r)$ has
specific energy $E(r)$ and specific angular momentum $L(r)$ given by
$$E(r) = {r\over2}{d\Phi \over dr} + \Phi~~{\rm and}~~
L(r) = \biggl(r^3 \ts{d\Phi \over dr}\biggr)^{1/2}~.$$
Then $dE/dL = \Omega(r)$, where $\Omega = (r^{-1} d\Phi/dr)^{1/2}$ is the
angular speed of rotation.  Disks spread when a unit mass at radius $r_2$
moves outward by acquiring angular momentum $dL$ from a unit mass at 
radius $r_1 < r_2$.  Is this energetically favorable?  The answer is yes.
The net change in energy, 
\begin{eqnarray}
dE = dE_1 + dE_2 & = & \biggl[- \biggl({dE \over dL}\biggr)_1 
                              + \biggl({dE \over dL}\biggr)_2\ts\biggr]dL\ts, 
                                                             \nonumber \\
                 &   &                                       \nonumber \\
                 & = & \ts[-\Omega(r_1) + \Omega(r_2)]\ts dL\ts, \nonumber
\end{eqnarray}
is negative because $\Omega(r)$ usually decreases outward.  ``Thus disk
spreading leads to a lower energy~state.  In general, disk spreading,
outward angular momentum flow, and energy dissipation accompany one another
in astrophysical disks'' (Tremaine 1989).  

\subsection{Self-Gravitating Systems Evolve By Spreading}

      The consequences are very general.  All of the following are caused by
the same basic physics.

      Globular and open clusters are supported by random motions, so they 
spread in three dimensions by outward energy transport.  The mechanism is
two-body relaxation, and the consequences are core collapse and the evaporation
of the outer parts.

      Stars are spherical systems supported by pressure.  They spread in three
dimensions by outward energy transport.  The mechanisms are radiation or
convection mediated by opacity.  Punctuated by phases of stability when nuclear
reactions replace the energy that is lost, stellar evolution consists of a
series of core contractions and envelope expansions.  One result is red
(super)giants.

      Protostars are spherical systems coupled to circumstellar disks.  The
spreading is complicated. The star shrinks in three dimensions; the inner
disk shrinks in two dimensions. Jets that look one-dimensional but that really
are three-dimensional carry away angular momentum (Shu et al.~1994, 1995).  
The mechanism involves magnetic fields wound up by differential rotation.  
Coupled to the outer circumstellar disk, they cause it to expand. 

      Protoplanetary disks are supported by rotation; they spread in two
dimensions by outward angular momentum transport.  Dynamical friction
produces, for example, hot Jupiters and colder Neptunes.

      Galactic disks are supported by rotation.  Exceptions are 
possible, but in general, they want to spread in
two dimensions by outward angular momentum transport.  Efficient driving
mechanisms are provided by bars and globally oval disks.
Like all of the above, the evolution is secular -- it is slow compared to
the collapse time of the disk.  Bar-driven secular evolution
is the subject of this paper.

\begin{figure*}[!t]
\includegraphics[width=9.5cm]{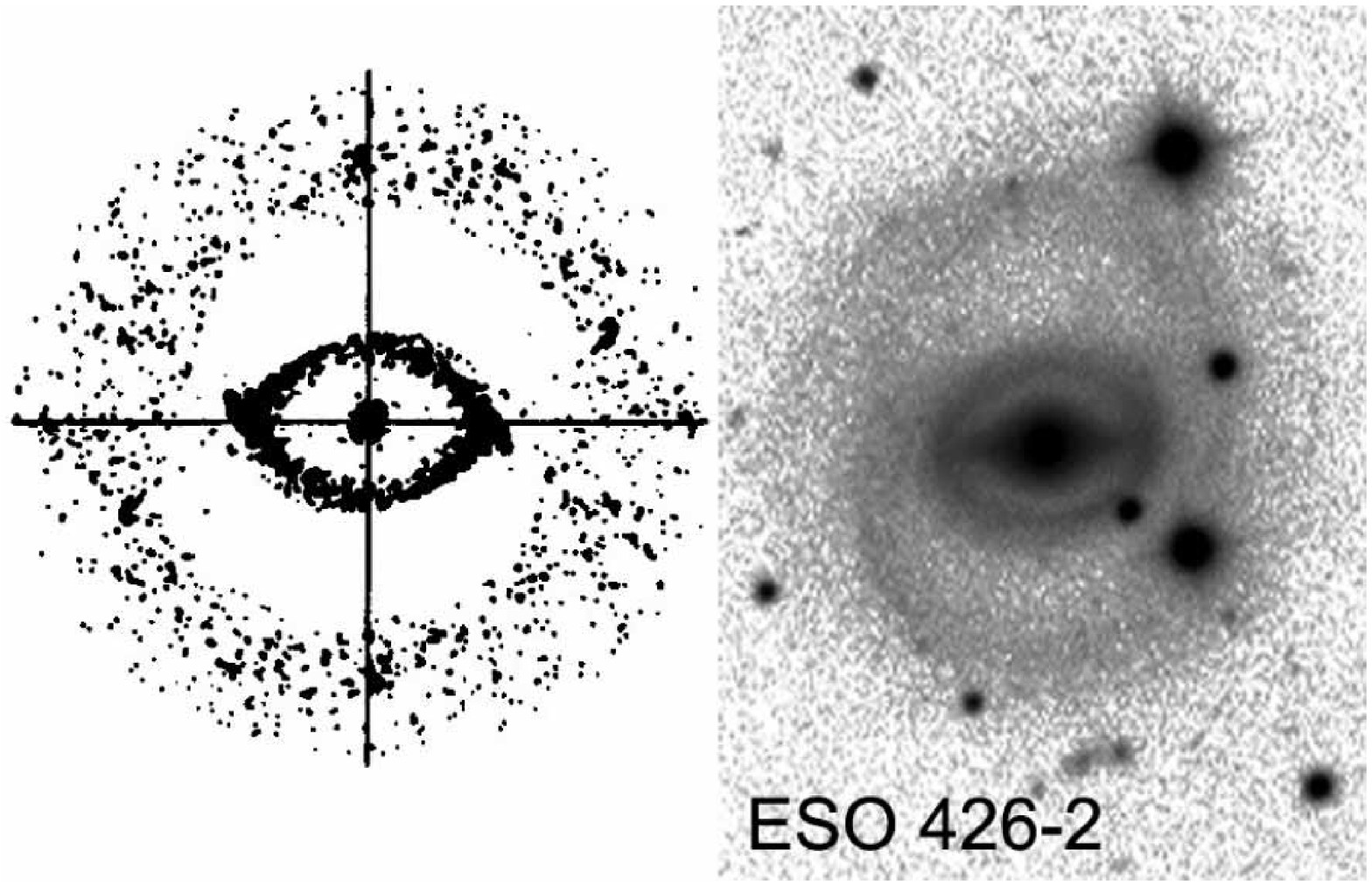}%
\hspace*{0.1cm}%
\includegraphics[width=7.3cm]{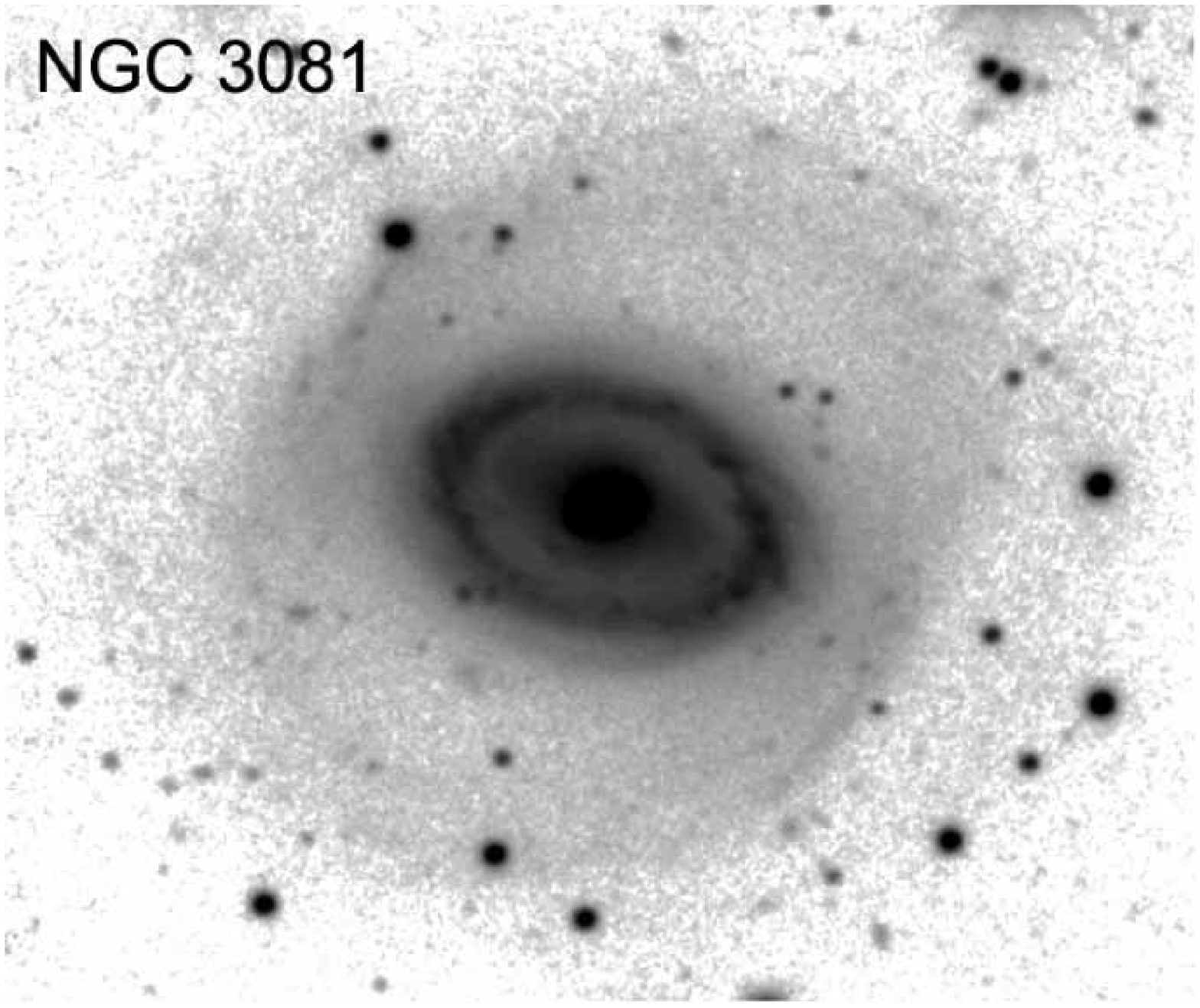}
\caption{Products of secular evolution: (left) Gas particle distribution
at the end of a sticky-particle simulation of the evolution of gas in a rotating
bar potential that is horizontal but not shown (Simkin, Su, \& Schwarz 1980).  After 
7 bar rotations, gas has collected into an outer ring, an inner ring around the
end of the bar, and a dense central concentration.  Similar features are seen 
in barred galaxies such as ESO 426-2 (Buta \& Crocker 1991) and NGC 3081 (Buta,
Corwin, \& Odewahn 2005).  This figure is adapted from Kormendy \& Kennicutt
(2004).}
\label{fig:dummy}
\end{figure*}

\section{How Bars Rearrange Disks}

      Many papers review simulations of bar-driven internal 
evolution (e.{\ts}g., Kormendy 1982a, 1993; Athanassoula 1992;
Sellwood \& Wilkinson 1993; Buta \& Combes 1996; KK04).  Interestingly,
simple sticky-particle
simulations reproduce the structure of barred galaxies better than 
hydrodynamic simulations that include more detailed physics, 
although the latter reveal important aspects of the evolution that 
the former cannot see.  Generic results are illustrated in Figure 1.
Disk gas is rearranged into an ``outer ring'' at $\sim 2.2$ bar radii,
an ``inner ring'' that encircles the end of the bar, and a dense central
concentration of gas.  As the gas density increases, star formation 
is likely, and indeed, the features produced in gas closely resemble 
the outer rings, inner rings, and (it will turn out) pseudobulges 
observed in stars in disk galaxies (see the figure).

      Morphological evidence consistent with the above interpretation
includes ring shapes and orientations.  Inner rings typically have 
axial ratios of $b/a \simeq 0.85$ and are oriented parallel to the
bar (Athanassoula et al.~1982; Buta \& Combes 1996).  Outer
rings have similar shapes and are oriented either parallel to or
perpendicular to the bar (Kormendy 1979; Simkin et al.~1980; Athanassoula
et al.~1982; Buta \& Combes 1996).  This is also consistent with the 
shapes of closed gas orbits if -- as expected -- bars typically end 
just inside corotation and if outer rings 
form just inside or just outside outer Lindblad resonance, respectively
(see the above papers).  Moreover, in galaxies of intermediate
Hubble types, in which bars are red and made of old stars while
disks are blue and dominated by young stars, the rings are also
blue and full of young stars.  Outer rings also generally 
contain gas and young stars, even in (R)SB0 galaxies. 
These points are illustrated in KK04.

      Rings are useful partly because they provide clean 
diagnostics of the evolution, but they contain only a small
fraction of the mass of the galaxy.  A bigger and ultimately
more important effect of the evolution results from the large
amount of gas that is driven toward the center by tidal
torques (e.{\ts}g., Athanassoula 1992).  In the 
simulations, it builds up to very high densities, often
in rings (see KK04 for a review).
Since star formation rate density $\Sigma_{\rm SFR}$ increases 
rapidly with gas density $\Sigma_{\rm gas}$, 
$\Sigma_{\rm SFR} \propto \Sigma_{\rm gas}^{1.4}$
(Kennicutt 1998a, b), high star formation rates are expected.
And indeed, they are observed.  KK04 review extensive
observations of nuclear starbursts, often in spectacular
rings and often associated with bars and globally
oval disks. They collect measurements of star formation rates
in starbursting nuclear rings and show that these extend
the above Kennicutt law from values seen in normal
galactic disks to high star formation rates and $\Sigma_{\rm gas}$ 
values.  With modest replenishment of the observed
nuclear gas, they would build stellar densities that
we observe in pseudobulges (see the next section) in
typically 1 -- 3 billion years.  So the formation
picture indicated by the simulations is plausibly 
connected via observed gas densities, star formation
rates, and reasonably timescales with the
disky bulges discussed in the next section.

\begin{figure*}[!ht]
\includegraphics[width=16.9cm]{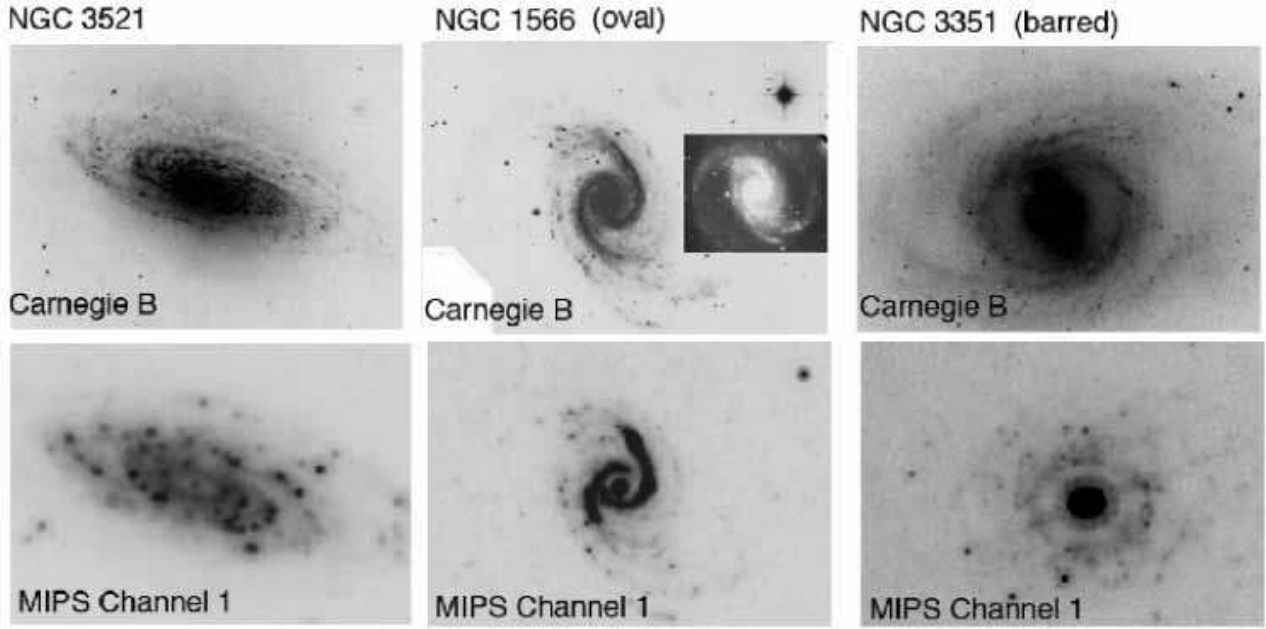}
\caption{Contrast the bright central 24 $\mu$m emission in NGC 1566,
a globally oval galaxy, and in NGC 3351, a barred galaxy, with the faint 
central 24 $\mu$m emission in NGC 3521, a galaxy that is neither barred nor
oval.  The 24 $\mu$m MIPS images are from the Spitzer Space Observatory SINGS
project (Kennicutt et al.~2003), while the comparison optical images are 
from the Carnegie Atlas of Galaxies (Sandage \& Bedke 1994).
}
\label{fig:dummy}
\end{figure*}

      Here, we illustrate the above picture with Spitzer 
Space Observatory images (Figure 2) of prototypical galaxies
that are neither barred nor oval (NGC 3521, left), globally oval
(NGC 1566, center), and barred (NGC 3351, right).  

      Ovals can be recognized kinematically (Bosma 1981) and 
photometrically (Kormendy \& Norman 1979; Kormendy 1979, 1982a; KK04). 
NGC 1566 has the photometric signature of a strong oval: it shows two 
distinct ``shelves'' in the brightness distribution with different 
axial ratios and position angles.  The high-surface-brightness, inner
shelf contains the most prominent spiral structure and is elongated 
N--S (vertically in Figure 2), while the outer shelf is much lower 
in surface brightness (see the inset) and is elongated horizontally 
in the figure.  Both cannot be round if they are coplanar, and studies
of H{\ts}I warps in edge-on galaxies show that the alternative -- 
warped disks --\ do not generally occur at such high surface 
brightnesses (Bosma 1981).  Each
nested oval generally has $b/a \sim 0.85$ (see the above
papers).  Ovals are important because they are easily 
nonaxisymmetric enough to drive secular evolution just
like that in  barred galaxies.

      The three galaxies shown in Figure 2 support the 
evolution picture discussed above.  The two galaxies that
have prominent disk nonaxisymmetries also are very bright
near the center in the Spitzer MIPS 24 $\mu$m images.
These are sensitive to warm dust that reradiates 
light from young stars.  That is, they indicate high
star formation rates.  In contrast, NGC 3521, which
has neither a bar nor an oval disk, shows little 24 $\mu$m 
emission near the center.  Three galaxies do not constitute 
a statistical sample, but optical observations suggest that 
the above behavior is typical (KK04).  Spitzer will provide 
quantitative checks of the statistics of such observations.

\section{Pseudobulge Properties}

     Kormendy (1982a, b) suggested that secular inward gas transport and star 
formation make disk-like ``bulges''.  
Combes \& Sanders (1981) suggested that boxy bulges formed from bars that 
heated themselves in the axial direction.  Pfenniger \& Norman (1990) discuss
both processes. A new dissipationless process -- that bars thicken in 
the axial direction as they decay -- is suggested by Klypin et al.~(2005).
These themes -- dissipational and dissipationless, secular pseudobulge building -- 
are widespread in the literature (see KK04 for review).

     How can we tell whether a ``bulge'' formed by these processes?
Fortunately, pseudobulges retain enough memory of their disky origin so that 
the best examples are recognizable.  Structural features that indicate a disky
origin are listed in KK04 and below.  We also give a few examples
and updates.

      Any prescription must recognize that we expect a continuum from
classical, merger-built bulges through objects with some E-like and some
disk-like characteristics to pseudobulges built completely by secular processes.
Uncertainties are inevitable when we deal with transition objects. Keeping these
in mind, a list of pseudobulge characteristics includes: 

\begin{enumerate}

\item The candidate pseudobulge is seen to~be~a~disk in images:~it has spiral
      structure or its ellipticity $\epsilon = 1 - b/a$ is similar to that of 
      the outer disk.\par
      One example among many, NGC 1353, is shown in Figure 3.  The images show, 
as Carollo et al.~(1997, 1998) concluded, that the central structure in NGC 1353
is a disk with 
similar flattening and orientation as the outer disk.  To make this quantitative, 
KK04 measured the surface brightness, ellipticity, and position angle profiles 
(plots in Figure 3).  The apparent flattening at $2^{\prime\prime}$ \lapprox
\ts$r$ \lapprox \ts$4^{\prime\prime}$ is the same as that of the main disk at
large radii.  The position angle is the same, too.  So the part of the galaxy
shown in the top-right panel really is a disk.  The brightness profile shows
that this nuclear disk is responsible for much of the central rise in surface
brightness above the inward extrapolation of an exponential fitted to the outer
disk.  Presented only with the brightness profile or with the bottom two panels
of images, we would identify the central rise in surface brightness as a
bulge.  Given Figure 3, we identify it as a pseudobulge.

\item It is or it contains a nuclear bar (in face-on galaxies).  Bars are disk
      phenomena; they are fundamentally different from triaxial ellipticals.

\item It is box-shaped (in edge-on galaxies).  Boxy bulges are believed to 
      be the central parts of edge-on bars that heated themselves in the axial 
      direction (see Sellwood \& Wilkinson 1993 for a review).  Again, bars
      are disk phenomena.

\item It has $n \simeq 1$ to 2 in a S\'ersic (1968) function, $I(r) 
      \propto e^{-K[(r/r_e)^{1/n} - 1]}$, fit to the brightness profile.~Here
      $n = 1$ for an exponential, \hbox{$n = 4$} for an
      $r^{1/4}$ law, and $K(n)$ is chosen so that radius $r_e$ contains half 
      of the light in the S\'ersic component.  We do not understand pseudobulge 
or disk formation well enough to predict $n$, so a nearly exponential profile
does not {\it prove} that a component is disk-like in the same way that high
flattening or large $V_{\rm max}/\sigma$ (see 5) do.  Instead,
combining $n$ with other pseudobulge indicators shows empirically that nearly
exponential profiles are a characteristic of many pseudobulges 
      (Andredakis \& Sanders 1994; Andredakis, Peletier, \& Balcells 1995;
      Courteau, de Jong, \& Broeils 1996; Carollo et al.~2002; 
      Balcells et al.~2003; MacArthur, Courteau, \& Holtzman 2003;
      KK04).  Again, NGC 1353 (Figure 3) is an example.  
KK04 decomposed the major-axis profile into an exponential outer disk
plus a S\'ersic function pseudobulge.  The best fit gave $n = 1.3 \pm 0.3$. \par
Whether this provides a classification criterion depends on whether classical
and pseudobulges are cleanly separated in correlations between $n$ and other 
parameters.  In this sense, a tentative result shown in Figure 4 is encouraging.
Several classical bulges for which high-accuracy $n$ values are 
available satisfy the $n$ -- $M_V$ correlation observed for elliptical galaxies
(e.{\ts}g., Caon et al.~1993; Graham \& Colless 1997; Fisher et al.~2005).
In our photometry, all of these have $n > 2$.  In contrast, all ``bulges'' that we
have studied and that appear to be pseudobulges based on other criteria have $n < 2$.
Work is in progress to enlarge the sample with very accurate $n$ values and to
further explore the above distinction (D.~Fisher's PhD thesis).

\item It is more rotation-dominated than are classical bulges in the
      \hbox{$V_{\rm max}/\sigma$ -- $\epsilon$} diagram; e.{\ts}g., 
      $V_{\rm max}/\sigma$ is larger than the value on the oblate line.

\item It is a low-$\sigma$ outlier in the Faber-Jackson (1976) correlation
      between (pseudo)bulge luminosity and velocity dispersion.  

\item It is dominated by Population I material (young stars, gas, and
dust), but there is no sign of a merger in progress. 

\end{enumerate}

      Small bulge-to-total luminosity ratios $B/T$ do not guarantee that a
galaxy contains a pseudobulge, but if $B/T$ \gapprox \ts1/2,
it seems safe to conclude that the galaxy contains a classical bulge.

      Based on these criteria, galaxies with classical bulges include M{\ts}31, 
NGC 3115, and NGC 4594.  NGC 1353 (Figure 3) and many galaxies illustrated in
KK04 contain prototypical pseudobulges.  The classification of the bulge of our 
Galaxy is ambiguous; the box-shaped structure favors a pseudobulge, but stellar
population data are most easily understood if the bulge is classical.

\vfill\eject
\clearpage

\begin{figure*}[!t]
\centerline{\includegraphics[width=14.8cm]{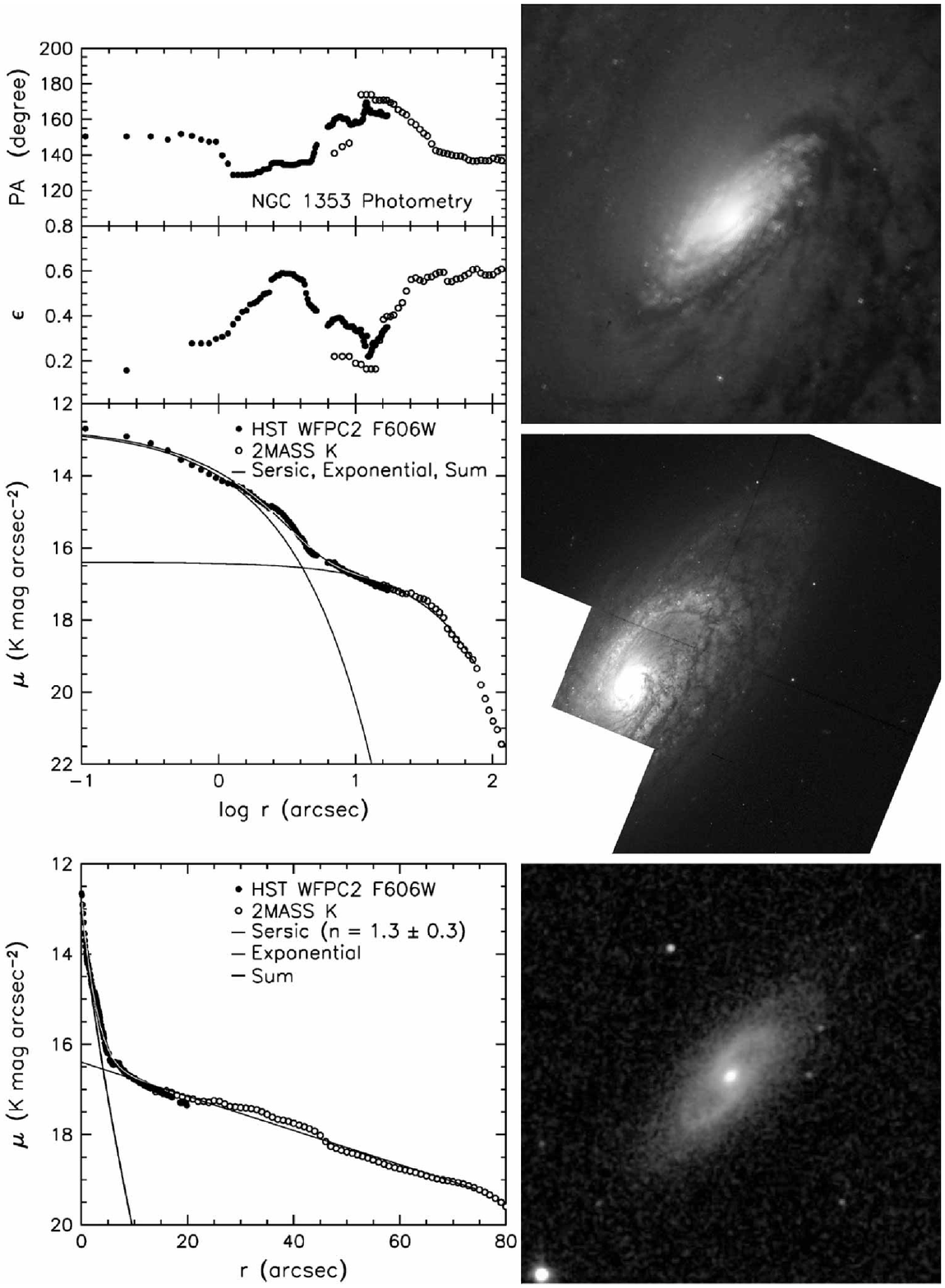}}
\caption{
\newdimen\sa  \def\sd{\sa=.1em \ifmmode $\rlap{.}$''$\kern -\sa$
                               \else \rlap{.}$''$\kern -\sa\fi}
\newdimen\sb  \def\md{\sb=.02em\ifmmode $\rlap{.}$'$\kern -\sb$
                               \else \rlap{.}$'$\kern -\sb\fi}
NGC 1353 pseudobulge (top image: 18$^{\prime\prime}$ $\times$
18$^{\prime\prime}$ zoom, and middle: full WFPC2 F606W image taken with 
{\it HST\/} by Carollo et al.~1998).  The bottom panel is a 2MASS (Jarrett et
al.~2003) $JHK$ composite image with a field of view of 4\md4 $\times$ 4\md4.
The plots show surface photometry with the {\it HST\/} profile shifted to the
$K$-band zeropoint.  The lines show a decomposition of the major-axis profile
into a S\'ersic (1968) function and an exponential disk.  The outer part of the
pseudobulge has the same apparent flattening as the disk.  This nuclear disk 
produces much of the rapid upturn in surface brightness toward the center. 
From Kormendy \& Kennicutt (2004).}
\label{fig:dummy}
\end{figure*}

\eject

\clearpage

\begin{figure}[!t]
\includegraphics[width=\columnwidth]{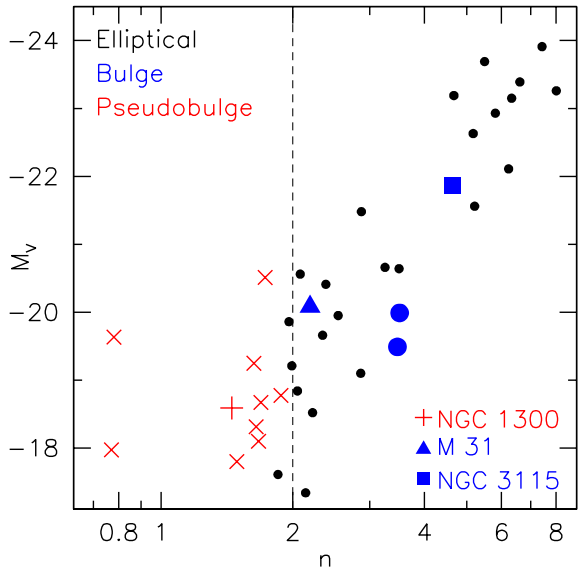}
\caption{Correlation of $n$ with bulge absolute magnitude $M_V$ for Virgo cluster
elliptical galaxies (small filled circles, from Fisher et al.~2005), 
classical bulges (large filled symbols) and pseudobulges (crosses and plus signs).
Three well known galaxies are identified.}
\label{fig:MabsNjk}
\end{figure}

\section{Given Hierarchical Clustering,\\
         How Can There Be So Many Bulgeless, Pure-Disk Galaxies?}

     Hierarchical clustering in a Universe dominated by cold dark matter 
(White \& Rees 1978) accounts remarkably well for large-scale structure but 
has trouble explaining the internal structure of galaxies.  One well known
problem is affected by secular pseudobulge formation.

      Hierarchical clustering produces merger violence.  {\it How could so many 
bulgeless, pure-disk galaxies form without undergoing major mergers} (T\'oth \& 
Ostriker 1992; Freeman 2000)?  Abadi et al.~(2003b) note that if satellites are 
accreted with suitable geometry, they can add to the disk, not the bulge.  This 
may help to explain old, thick disks (Mould 2005).  But it helps only after 
the galaxy has become big enough so that inhaling a satellite is a minor 
accretion that results in satellite stripping, not violent relaxation 
that disrupts disks.~By the time a galaxy gets this big, hierarchical 
clustering generally gives it a bulge (Steinmetz \& Navarro 2002; Abadi 
et al.~2003a,{\ts}b; Meza et al.~2003).  Baryonic physics helps (reionization,
supernova-driven energy feedback:~Navarro, Eke, \&  Frenk 1996;
Moore et al. 1999; Klypin et al.~1999).~But this problem is hard.  

      We want to know how hard it is. Schechter \& Dressler (1987) and Benson 
et al.~(2002) estimate that approximately equal amounts of mass are incorporated into
bulges and disks.  This is based on photometric decompositions into $r^{1/4}$-law 
``bulges'' and exponential disks.

      But if secular evolution turns some disk material into pseudobulges that
get confused with classical bulges in the above decompositions, then we
overestimate the mass in classical bulges by a modest amount (KK04).   The 
bulge-disk decompositions of Simien \& de Vaucouleurs (1986) already suggest 
that late-type galaxies have very small bulges.  Sbc, Sc, Scd, and Sd galaxies 
are found to have median bulge-to-total luminosity ratios of 0.17, 0.1, 0.03, 
and 0.02, respectively.  When people try to understand this result in the context
of hierarchical clustering, they hope that the gentlest part of the distribution
of formation histories produces only a small merger contribution compared to
dominant quiescent accretion.  Still:~``Reconciling \dots~the properties of
disk galaxies with the \dots~high merging rates characteristic of hierarchical
formation scenarios such as $\Lambda$CDM remains a challenging, yet so far 
elusive, proposition'' (Abadi et al.~2003a).

      The problem is worse for the numbers of classical bulges.   We now know 
(KK04) that there is a sharp transition between Sb and Sc; Sb- and earlier-type 
galaxies mostly contain classical bulges; Sbc galaxies contain pseudobulges more 
often than bulges, and Sc- and later-type galaxies appear never to have classical
bulges.  This implies that a majority of field galaxies -- not just a few, 
remarkably flat, edge-on, bulgeless disks (Matthews, Gallagher, \& van Driel 
1999; Freeman 2000; van der Kruit et al.~2001) -- show no signs of major 
merger violence. The challenge for hierarchical clustering is correspondingly
increased. 

      We emphasize:~we are not trying to disprove hierarchical clustering.  
But we do see signs that pieces of the evolution puzzle are missing.  Our aim 
is to put the observational challenge on as quantitative a footing as possible.
We need a better determination of the luminosity functions of classical and 
pseudo bulges.  This will lead to a better understanding of how far galaxies can evolve 
toward earlier Hubble types as a result of secular processes.

\acknowledgements

We are grateful to Scott Tremaine for helpful comments on the MS and for 
permission to reproduce his discussion in Section 2.2.  We also thank 
Ron Buta and Marcella Carollo for providing many of the images used in
the figures.  Mark Cornell collaborates with us on the McDonald Observatory
0.8 m telescope imaging and on many aspects of pseudobulge research.  This 
paper is based partly on observations made with the NASA/ESA {\itshape 
Hubble Space Telescope}, obtained from the data archive at the Space
Telescope Science Institute.  STScI is operated by AURA, Inc. under NASA 
contract NAS 5-26555.  We also used the NASA/IPAC Extragalactic Database
(NED), which is operated by JPL and Caltech under contract with NASA.

\end{document}